\documentclass[aps,prl,twocolumn,nofootinbib,groupedaddress]{revtex4}
\usepackage{graphicx}

\begin{document}


\title{Dark Energy as a Born-Infeld Gauge Interaction
Violating the Equivalence Principle}


\author{A. F\"uzfa, J.-M. Alimi}
\affiliation{Laboratory Universe and Theories, CNRS UMR 8102,\\
 Observatoire de Paris-Meudon and Universit\'e Paris VII, France}



\begin{abstract}
We investigate the possibility that dark energy does not couple to gravitation in the same way as ordinary matter, 
yielding a violation of the weak and strong equivalence principles on cosmological scales. We build a transient mechanism 
in which gravitation is pushed away from general relativity by a Born-Infeld gauge interaction acting as an 
"\textit{Abnormally Weighting (dark) Energy (AWE)}". This mechanism accounts for the Hubble diagram of far-away supernovae by cosmic acceleration and time variation 
of the gravitational constant while accounting naturally for the present tests on general relativity.
\end{abstract}


\maketitle

To account for the dimmed magnitude of type Ia supernovae (see \cite{snls} and references therein), it is
necessary to invoke a recent acceleration of the cosmic expansion,
provided these objects can be considered as standard
candles. This usual explanation does not require
to give up general relativity (GR) as it includes naturally a way to accelerate
cosmic expansion through a positive cosmological constant $\Lambda$. In the
standard cosmological picture, based on GR, gravitation contains
only spin 2 gravitational degrees of freedom (the metric field
$g_{\mu\nu}$) and obeys the equivalence principle. Under the assumptions of the cosmological
principle, the corresponding geometry for space-time is locally
given by the Friedmann-Lemaitre-Robertson-Walker (FLRW) line
element:
\begin{equation}
\label{flrw}
ds^2=-dt^2+a^2(t)\left(dr^2+r^2d\theta^2+r^2\sin^2\theta
d\varphi^2\right)
\end{equation}
where $a(t)$ is the scale factor and where we assumed synchronous
time coordinate\footnote{Here, we have also restricted ourselves to the case
of flat space-times for the sake of simplicity. Throughout this paper, we will assume the Planck
system of units, in which $\hbar=c=1$, $G=m_{Pl}^{-2}$ and the gravitational
coupling constant is $\kappa=8\pi G$.}. In this framework,
the cosmic acceleration is ruled by the following equation:
\begin{equation}
\label{acc1} \frac{\ddot{a}}{a}=-\frac{4\pi}{3} G \left(\rho+3p\right),
\end{equation}
where $\rho$ and $p$ stand for the energy density and pressure
of the matter filling space-time. In order to provide the necessary
cosmic acceleration ($\ddot{a}>0$), it is therefore compulsory to
violate the strong energy condition (SEC) \cite{caroll}: $p<-\rho/3$. 
The Einstein cosmological constant $\Lambda$ is the usual way used to provide 
cosmic acceleration, although it leads to the intricate problems of fine-tuning ($\rho_\Lambda^{th}\approx m_{Pl}^4\approx 10^{76}GeV^4$)
and coincidence ($\rho_\Lambda^{obs}\approx\rho_{c,0}=3H_0^2/(8\pi G)\approx 10^{-47}GeV^4$) once $\Lambda$ is interpreted as non-vanishing
vacuum energy density (cf. \cite{weinberg} for a review and \cite{padmanabhan} for an interesting alternate interpretation). 
Most of the alternate explanations, like quintessence, also require to violate the SEC.

In this letter, we propose a completely new interpretation of dark energy that does not require 
this violation. Instead, we assume that "dark" energy violates the weak
equivalence principle (WEP) on large-scales, i.e. it does not couple to
gravitation as usual matter and weights abnormally. Doing so, its related gravitational binding
energy will be felt differently by other types of matter, therefore
violating also the strong equivalence principle (SEP). Under these
assumptions, we build a dark energy mechanism without violation of the SEC. 
This \textit{Abnormally Weighting
Energy} (AWE) will consist here of an additional gauge interaction of
Born-Infeld (BI) type which will provide a natural scheme for transient
dark energy mechanism. This will lead to a
satisfactory explanation of Hubble diagrams of type Ia supernovae
while still accounting for the stringent constraints on GR
we know today.

Here, we will consider that the energy
content of the universe is divided into three parts : a
gravitational sector described by pure spin 2 (graviton) and spin 0
(dilaton) degrees of freedom, a matter sector containing the usual
fluids of cosmology (baryons, photons, dark matter, ...)
and an AWE sector, here composed by a gauge interaction ruled
by BI type gauge dynamics. The introduction of a
scalar partner to the graviton is necessary to account for the violation
of the equivalence principle. The violation of the WEP by the AWE 
can be represented by different couplings between gravity,
the AWE and usual matter:
\begin{eqnarray}
\label{action1}
S&=&\frac{1}{2\kappa}\int\sqrt{-g}d^4x\left\{R-2g^{\mu\nu}\partial_\mu\varphi\partial_\nu\varphi\right\}\nonumber\\
&& +S_{BI}\left[A_\mu,A_{BI}^2(\varphi)g_{\mu\nu}\right]\nonumber\\
&&+S_m\left[\psi_m,A_m^2(\varphi)g_{\mu\nu}\right],
\end{eqnarray}
where $\kappa$ is the "\textit{bare}" gravitational coupling
constant.  In the previous action, $g_{\mu\nu}$ is the Einstein metric,
$\varphi$ is a gravitational scalar field, $S_{BI}$ is a gauge AWE
sector ruled by BI dynamics and $S_m$ is the usual matter
sector with matter fields $\psi_m$; $A_{BI}(\varphi)$ and $A_m(\varphi)$ being the coupling functions 
to the metric $g_{\mu\nu}$ for the AWE and matter sectors respectively. The non-universality of the
gravitational couplings ($A_{BI}\ne A_m$) yields a violation of the
WEP: experiments using the new
BI gauge interaction would provide a different inertial mass
than all other experiments\footnote{Furthermore, non-universal couplings to the dilaton arise naturally in 
string theory, see \cite{damour} for example.}. The action (\ref{action1}) is written
in the so-called "\textit{Einstein frame}" where the metric
components are measured by using purely gravitational rods and
clocks, i.e. not build upon any of the matter fields nor the ones from the AWE sector. 
We will define the "\textit{Dicke-Jordan}" observable frame
by the conformal transformation
\begin{equation}
\label{obsframe} \tilde{g}_{\mu\nu}=A^2_m(\varphi)g_{\mu\nu}
\end{equation}
using the coupling function to ordinary matter. Indeed, in this
frame, the metric $\tilde{g}_{\mu\nu}$ couples universally to
ordinary matter and is measured by clocks and rods made of usual
matter (but not build upon the new gauge interaction we introduced
as the AWE sector). The violation
of the WEP therefore only concerns the new
gauge sector that was introduced in (\ref{action1}). Throughout this
paper, quantities with a tilde will refer to the observable frame
given
by (\ref{obsframe}).

BI gauge dynamics allows to avoid point-like singularities  
in the field strength
through classical vacuum polarization effects by freezing the gauge potentials 
above some given critical energy $\epsilon_c$. This can
be done by assuming the lagrangian 
$\mathcal{L}_{BI}=\epsilon_{c}\left(\mathcal{R}-1\right)$ for the gauge field, where 
$$
\mathcal{R}=\sqrt{1+A_{BI}^{-4}/\left(2\epsilon_{c}\right)F_{\mu\nu}F^{\mu\nu}-A_{BI}^{-8}/\left(16\epsilon_c^2\right)\left(F_{\mu\nu}\tilde{F}^{\mu\nu}\right)^2}
$$
(see \cite{dyadichev} and references therein).
At low-energies, the BI gauge
dynamics reduces to Yang-Mills dynamics where the gauge fields are radiative. 
In a cosmological context (see \cite{dyadichev} for a complete study of cosmology with BI gauge fields
and \cite{fuzfa2} for the introduction of the dilaton $\varphi$ in the model), such BI gauge fields obey 
the following equation of state:
\begin{equation}
\label{eos}
\omega_{BI}=\frac{p_{BI}}{\rho_{BI}}=\frac{1}{3}\left(\frac{\epsilon_{c}-A_{BI}^{-4}(\varphi)\rho_{BI}}{\epsilon_{c}+A_{BI}^{-4}(\varphi)\rho_{BI}}\right),
\end{equation}
where $\rho_{BI}$ ($p_{BI}$) is the gauge energy density (pressure) in the Einstein frame. 
As the only coupling between AWE and matter is purely gravitational, 
the scaling evolution of the gauge energy density is
$$
\rho_{BI}=\epsilon_{c}A_{BI}^{4}(\varphi)\left(\sqrt{1+\mathcal{C}/(A_{BI}^{4}(\varphi)a^4})-1\right)
$$
($\mathcal{C}$ is an integration constant). When the
condition $A_{BI}^{-4}(\varphi)\rho_{BI}\gg\epsilon_c$ occurs, the
gauge field pressure is negative $p_{BI}/\rho_{BI}\approx -1/3$, 
and the related gauge field
energy density scales as $\left(A_{BI}(\varphi)a\right)^{-2}$ (frozen field strength). 
However, in the low-energy regime\\
$A_{BI}^{-4}(\varphi)\rho_{BI}\ll\epsilon_c$, the fluid becomes
relativistic $p_{BI}/\rho_{BI}\approx 1/3$. This remarkable equation of state 
allows a possible transient domination of the BI energy.

The general cosmological dynamics of the action (\ref{action1}) have
been studied in details in \cite{fuzfa2} for various couplings of
the gauge field to the dilaton, while the case of BI gauge fields alone can be found already
in \cite{dyadichev}. Here, we will focus on describing
the transient dark energy mechanism based on this dynamics.
The Friedmann equation obtained from the action (\ref{action1}) writes down
\begin{equation}
\label{hebid}
\left(\frac{\dot{a}}{a}\right)^2=\frac{\dot{\varphi}^2}{3}+\frac{\kappa}{3}\left[\rho_{BI}+\rho_m\right],
\end{equation}
where a dot denotes a derivative with respect to the time coordinate
$t$, and $\rho_m$ is
the energy density of the matter sector (Einstein frame). The acceleration
equation is given by
\begin{equation}
\label{acc}
\frac{\ddot{a}}{a}=-\frac{2}{3}\dot{\varphi}^2-\frac{\kappa}{6}\left[\left(\rho_{BI}+3p_{BI}\right)+\left(\rho_m+3p_m\right)\right]\cdot
\end{equation}
There cannot be any cosmic
acceleration in terms of the metric $g_{\mu\nu}$ (the dilaton
$\varphi$ is here massless), as the highest value
of $\ddot{a}$ that can be achieved in this frame is identically zero
from (\ref{acc}) (see \cite{fuzfa2}), as the BI gauge interaction
never violates the SEC. As there is no direct coupling between the gauge and the matter
sectors in (\ref{action1}), the behavior of the matter energy density and pressure are
given as in usual tensor-scalar gravity. These quantities are given
in the observable frame (\ref{obsframe}) by
$\tilde{\rho_m}=A_{m}^{-4}(\varphi)\rho_m$ where $\rho_m$ represents these quantities expressed in the
Einstein frame (with similar relation for the pressure). 
In this frame, they have the same scaling law as in standard cosmology based on GR. 

The scalar gravitational dynamics are given by the Klein-Gordon
equation:
\begin{eqnarray}
\label{kg}
\ddot{\varphi}+3\frac{\dot{a}}{a}\dot{\varphi}+&\frac{\kappa}{2}\alpha_{BI}(\varphi)\left(\rho_{BI}-3p_{BI}\right)&\nonumber\\
&+\frac{\kappa}{2}\alpha_m(\varphi)\left(\rho_m-3p_m\right)&=0,
\end{eqnarray}
where $\alpha_i(\varphi)=d\ln A_i(\varphi)/d\varphi\cdot$
The violation of the WEP induced by the AWE sector ($\alpha_{BI}\ne\alpha_m$) 
implies that the history of the universe can be seen
as a competition between usual matter and AWE, particularly if the first
attracts the field toward values corresponding to GR (here $\varphi=0$ and $\dot{\varphi}=0$) 
while the last acts as a repulsion from it.
As the AWE sector is here constituted by a BI gauge interaction, this competition will be temporary because of
the equation of state (\ref{eos}).
At high-energies, the negative pressure will first
allow a late domination of AWE, while at low-energies the radiation behavior ($\omega_{BI}=1/3$) will ensure both
a decoupling of the AWE 
sector from the scalar field (see (\ref{kg})) and a final sub-dominance of AWE.
The resulting dark energy mechanism is therefore transient. 
A well-known and remarkable feature of tensor-scalar theories of gravitation is
their convergence towards general relativity during
matter-dominated era (see \cite{alimi} and references therein), which is ensured
when the coupling function $\alpha_m(\varphi)$ has a global minimum or which can be achieved provided
specific initial conditions for general coupling functions. 
In order to introduce a competition between attraction by ordinary matter
and repulsion by AWE in (\ref{kg}), it suffices to assume the usual coupling functions:
$A_{BI}(\varphi)= \exp\left(k_{BI}\varphi\right)$ and $A_{m}(\varphi)= \exp\left(k_{m}\frac{\varphi^2}{2}\right)\cdot$
The deviation from GR occurs when the dilaton
$\varphi$ is pushed away from the minimum of the matter coupling
function $A_m(\varphi)$ (GR) by the constant drag term ($\alpha_{BI}=k_{BI}$) when
the BI term in (\ref{kg}),
$\alpha_{BI}(\varphi)\left(\rho_{BI}-3p_{BI}\right)\gg \alpha_{m}(\varphi)\rho_{m}$,
dominates the scalar dynamics (matter-dominated era $p_m\approx 0$).
However, convergence to GR is ensured by the efficiency
of the attraction mechanism associated to the coupling function
$\alpha_{m}=k_{m}\varphi$, provided the matter force term in (\ref{kg}) dominates, which occurs twice.
The first time is before the AWE dominance, when the BI gauge interaction was sub-dominant, and this
allows to account for the validity of GR in the early times
of Cosmic Microwave Background (CMB) and Big Bang
Nucleosynthesis (BBN). The second time is at the end of the process
as soon as the BI gauge interaction behaves like radiation, i.e. even if it is still dominating the energy content of the universe. 

Let us now illustrate this mechanism, where
dark AWE never violates the
SEC $p>-\rho/3$ in the Einstein frame, by reproducing
a Hubble diagram built upon recent available data on far-away type Ia
supernovae \cite{snls}. Within the framework of tensor-scalar
gravity, the dimmed magnitude of such objects could be explained
both by an acceleration of cosmic expansion and a time variation of
the gravitational constant. In \cite{gaztanaga,riazuelo}, the following
toy model for the moduli distance vs redshift relation of type Ia
supernovae has been proposed:
\begin{equation}
\label{muz} \mu(\tilde{z})=m-M=5\log_{10} d_L(\tilde{z})
+\frac{15}{4} \log_{10} \frac{G_{eff}(\tilde{z})}{G_0},
\end{equation}
where $d_L(\tilde{z})$ is the luminous distance
(in Mpc) given by $d_L(\tilde{z})=(1+\tilde{z})\tilde{H}_0\int_0^{\tilde{z}}
d\tilde{z}/\tilde{H}(\tilde{z})$
for a flat universe ($\tilde{H}_0$ is the observed value of the
Hubble constant today). The expansion rate $\tilde{H}(\tilde{z})$
has to be estimated in the observable frame related to usual matter
(\ref{obsframe}) ($\tilde{H}=d\tilde{a}/(\tilde{a}d\tilde{t})=A_m^{-1}(\varphi)\left(H+\alpha_m(\varphi)\dot{\varphi}\right)$,
with $H=\dot{a}/a$ is the Hubble parameter in the Einstein frame).
In (\ref{muz}), $G_{eff}$ is
the effective gravitational "\textit{constant}" at the epoch
$\tilde{z}$:
\begin{equation}
\label{gn} G_{N}=G_0 A^{2}_m(\varphi)(1+\alpha_m^2(\varphi))\cdot
\end{equation}
where $G_0$ is the (bare) value of this constant today,
where gravitation is well-described by GR. 
In addition to account for moduli distance data, any dark energy
mechanism based on tensor-scalar theory of gravitation should be in
agreement with the present tests of GR, which concern only usual matter and not AWE. The
constraints on post-newtonian parameters are given by (cf. \cite{bertotti,will}):
\begin{eqnarray}
\label{ppn1} |\gamma-1|&=&2
\frac{\alpha_m^2(\varphi)}{\left(1+\alpha_m^2(\varphi)\right)}<2\times 10^{-5}\\
|\beta-1|&=&|\frac{d\alpha_m}{d\varphi}\frac{\alpha_m^2(\varphi)}{2\left(1+\alpha_m^2(\varphi)\right)^{2}}|<6\times
10^{-4}\cdot\label{ppn2}
\end{eqnarray}
Another constraint is the time-variation of the gravitational
constant \cite{dickey}:
\begin{eqnarray}
\label{ppn3} |\frac{\dot{G}}{G}|&=&
|2\dot{\varphi}\alpha_m(\varphi)\frac{1+\frac{d\alpha_m}{d\varphi}}{1+\alpha_m^2(\varphi)}|<6\times
10^{-12}yr^{-1}\cdot
\end{eqnarray}
One should also add the constraints on the WEP, 
which is tested at the $10^{-12}$ level by the
universality of free fall of inertial masses with different
compositions \cite{damour4}. Although the AWE violates this universality of free fall, we
might consider that this effect is extremely weak (and not observed
in practice) provided the AWE density at our scale is of the
order of its cosmological value ($\rho_{BI,0}\approx\rho_{c,0}$). 
This is true if the
AWE does not cluster too much at our scale, an assumption
that should be verified in forthcoming works. Therefore, we will
only consider the constraints (\ref{ppn1}), (\ref{ppn2}) and
(\ref{ppn3}) while discarding the effects on the universality of
free fall for the moment.

\begin{figure}
\begin{center}
\includegraphics[scale=0.3]{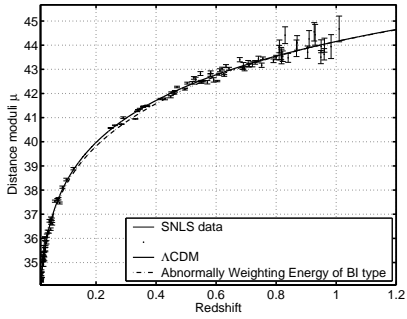}
\end{center}
\caption{Hubble diagram of SNLS 1st year data set with the best fit
$\Lambda CDM$ flat model (solid line, $\Omega_m(a_0)=0.26$,
$\bar{\chi}^2/dof=1.03$) and the AWE model (dash-dotted line,
$\bar{\chi}^2/dof=1.09$) ($H_0=70 km/s/Mpc$) } \label{fig1}
\end{figure}

The dark energy model proposed here actually depends
on four free parameters: the initial BI energy density $\tilde{\rho}_{BI}(a_i)$,
the critical BI energy
$\epsilon_c$, the parameters $k_{BI}$ and $k_m$ of the
two dilaton couplings to the AWE and to matter,
respectively. Once the cosmological evolution is determined (see \cite{fuzfa2} for details), 
the observable quantities are derived using
(\ref{obsframe}) and (\ref{gn}). 
The parameters $\tilde{\rho}_{BI}(a_i)$ and $\epsilon_c$
are chosen such that $\tilde{\Omega}_m(a_0)\approx
0.3$ (flat universe), the observable energy contributions being given by performing 
the conformal transformation (\ref{obsframe}) on
the Friedmann equation (\ref{hebid}).
Figure \ref{fig1} illustrates the adequacy of the model to
a Hubble diagram of type Ia supernovae. As a matter of comparison, we also give
the value of the $\chi-$square, marginalised over $H_0$, per degrees
of freedom denoted by $\bar{\chi}^2/dof$. The model
was first set by minimising the $\bar{\chi}^2$ to a value
very close to the best fit $\Lambda CDM$
model. Then, as the constraints (\ref{ppn1}), (\ref{ppn2}) and
(\ref{ppn3}) were not completely satisfied for these best fit models,
we pushed the time-integration a little bit further to let the
attraction mechanism fix this naturally. This resulted in a slightly increased value
of $\bar{\chi}^2/dof$.
We will not go deeper here into these statistical issues, as our aim
is only to illustrate the dark energy mechanism described here. 

\begin{figure}
\begin{center}
\begin{tabular}{c}
\includegraphics[scale=0.3]{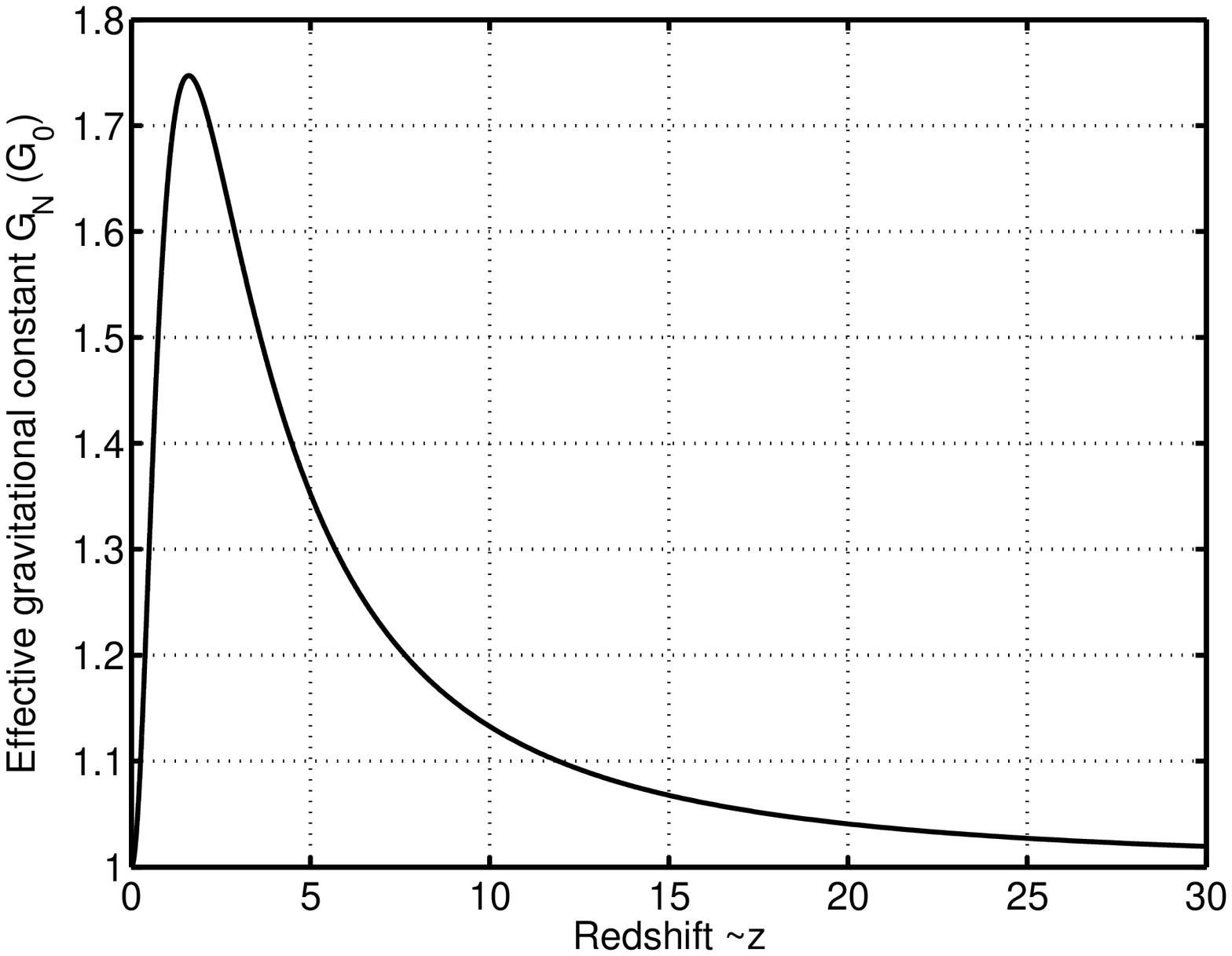} \\
\includegraphics[scale=0.3]{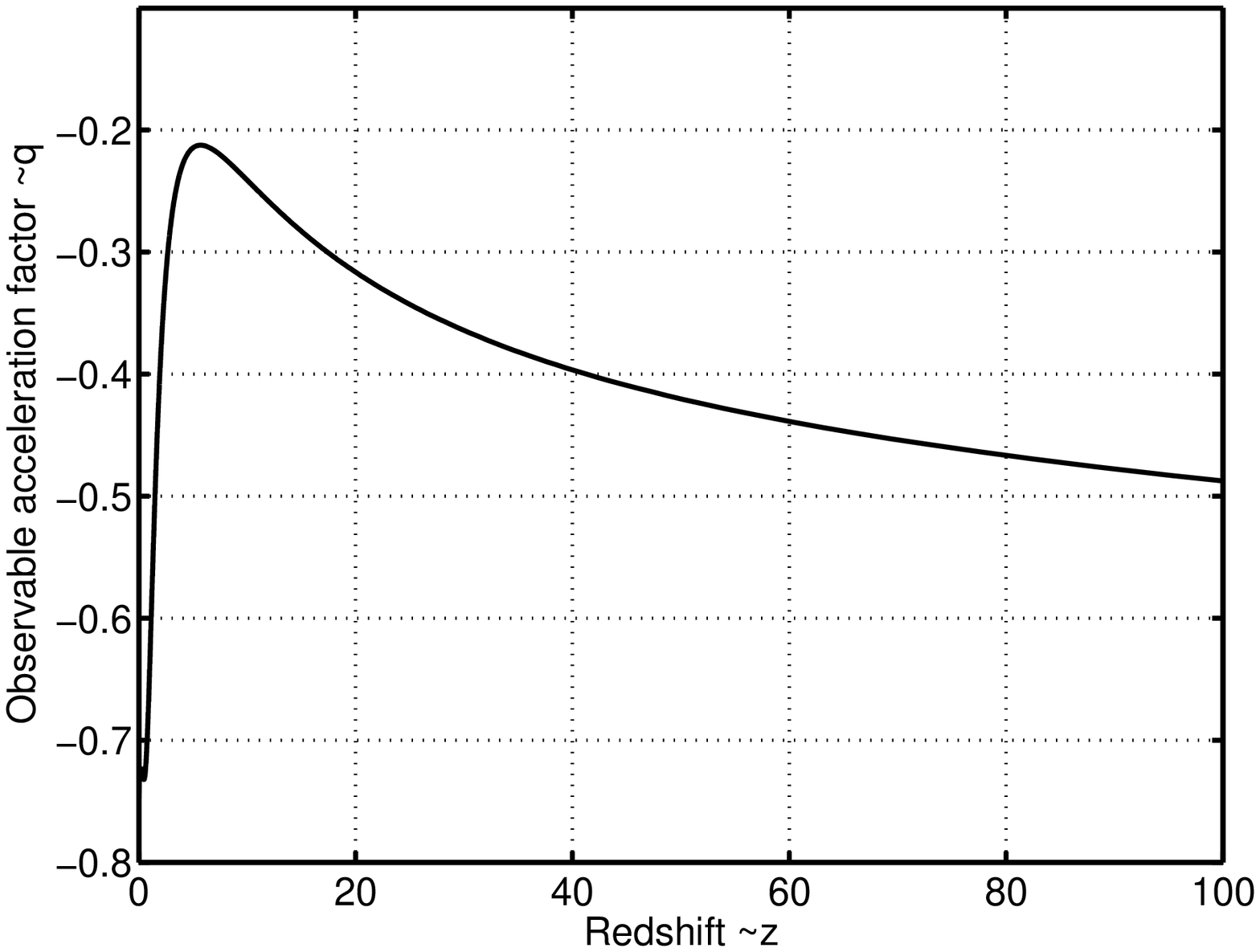}
\end{tabular}
\end{center}
\caption{a) Cosmological evolution of $G_N$ (in units of
bare $G$) b) Evolution of the acceleration factor $\tilde{q}=1+\left(d\tilde{H}/d\tilde{t}\right)/\left(\tilde{H}^2\right)$ with the redshift 
$\tilde{z}$ in the observable frame} \label{fig2}
\end{figure}

Figure \ref{fig2}a) represents the corresponding cosmological evolution of
the effective gravitational constant (\ref{gn}) while Figure
\ref{fig2}b) illustrates the cosmic acceleration. For the  
correction due to a variable $G_N$ in the toy model (\ref{muz}), the model
does not require any cosmic acceleration
(see Figure \ref{fig2}b)). However, we have found that if the correction in $G_N$ 
has been over-estimated in (\ref{muz}), transient cosmic acceleration is needed to match data. 
This acceleration is only due to the
interpretation in the observable frame (\ref{obsframe}) and not to a violation of the SEC 
in the Einstein frame (see also \cite{fuzfa2}). 
Therefore, dark energy effects consist here of a combination of variable $G_N$ 
and transient cosmic acceleration. 
Figure 3 represents the evolution of the post-newtonian parameters
(\ref{ppn1}), (\ref{ppn2}) and the constraints on the absolute
variation of $G_N$ (\ref{ppn3}). The convergence occurs 
during domination of the AWE sector because of the decoupling of the BI gauge
interaction once it reaches its radiative regime.
\begin{figure}
\begin{center}
\includegraphics[scale=0.3]{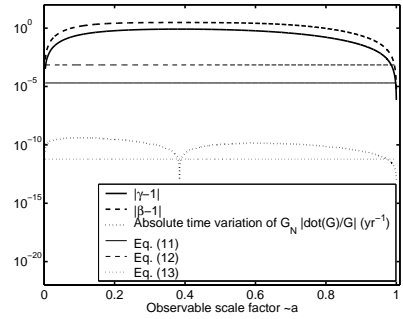}
\end{center}
\caption{Evolution of post-newtonian parameters with the scale factor
$|\gamma-1|$ (solid line); b) $|\beta-1|$ (dashed line) ; c) $|\dot{G}/G|$ (dots). Current observable
constraints are indicated by the horizontal lines} \label{fig3}
\end{figure}
The history of the mechanism is as follows : 
the BI gauge field starts sub-dominant at
the end of the radiation-dominated era while gravitation is
well described by GR. Then, as the energy densities progressively cool down to coincidence
the scalar field is pushed away from GR by the increasing
repulsive influence of the AWE. This repulsive
influence rapidly decreases
as the gauge field becomes radiative and
decouples from the scalar sector. Between this period and today,
matter becomes the dominant driving term and
attracts towards GR to finally achieve the level of precision we
know for it today. However, during a short period of time
in the very recent cosmic history, gravitation was substantially
different from GR and led to dark energy effects.

This letter has presented a new dark energy mechanism where 
this energy weighs abnormally. This violates the
WEP which obviously leads to a violation of
the SEP as modeled by a tensor-scalar
theory of gravitation. As a consequence, the "dark" AWE does not need anymore to exert too
negative pressures (and a violation of the SEC)
to achieve its job efficiently.
The BI gauge interaction used as AWE provides here a natural
transient dark energy mechanism compatible with
supernovae data, constraints on GR today and during the radiative era. 
However, this mechanism is likely to have a strong impact on physics in
the matter-dominated era by the variation of $G_N$ and the acceleration it yields. 
As well, the
clustering of such AWE should also lead to a
violation of the universality of free fall. A careful study of
all these effects could therefore determine whether some processes in
the universe are not ruled by the equivalence principle. 
If proved true, this would completely
change our views of gravitation and the universe.

\end{document}